\documentclass[aps,prl,preprint,superscriptaddress]{revtex4-1}
\usepackage{graphicx}
\usepackage{hyperref}

\hypersetup{
	colorlinks=true,
	linkcolor=blue,
	citecolor=blue,
	urlcolor=blue
}

\newcommand{\omcavi}{\omega_{l}} 			
\newcommand{\omcavii}{\omega_{c}} 			
\newcommand{\ncavii}{n_c}					
\newcommand{\ncavi}{n_l}					
\newcommand{\vcav}{V_c}						
\newcommand{\wodi}{\textrm{WD}} 			
\newcommand{\om}{\Omega_m} 					
\newcommand{\omx}{\Omega_x}
\newcommand{\omy}{\Omega_y}
\newcommand{\omz}{\Omega_z}
\newcommand{\kb}{k_{B}} 					
\newcommand{\garc}{\Gamma_{rc}} 			
\newcommand{\xzpf}{x_{zpf}}					

\newcommand{\exCq}{$C_Q=0.01$} 						
\newcommand{\exGammaCq}{$\Gamma=2\pi\times175$ kHz}	
\newcommand{\exKappa}{$\kappa=2\pi\times193$ kHz}		
\newcommand{\exCoupling}{$g=2\pi\times 9.6$ kHz}			
\newcommand{\exBasePressure}{$p=4.8\times 10^{-7}$ mbar} 	
\newcommand{\exPressureOmit}{$p=4.3\times10^{-6}$ mbar}	
\newcommand{\exCavityWaist}{$w_0=41.1\ \mu$m} 			
\newcommand{\exCavityModeMirror}{$w=61\ \mu$m}			
\newcommand{\exWaistY}{$w_y=(42.3\pm1.2)\ \mu$m}		
\newcommand{\exWaistZ}{$w_z=(41.8\pm1.2)\ \mu$m}		

\begin{document}

\title{Levitated cavity optomechanics in high vacuum}	

\author{Uro\v s Deli\' c}
\thanks{U.D. and D.G. contributed equally to this work.}
\affiliation{Vienna Center for Quantum Science and Technology (VCQ), Faculty of Physics, University of Vienna, Boltzmanngasse 5, A-1090 Vienna, Austria}

\affiliation{Institute for Quantum Optics and Quantum Information (IQOQI), Boltzmanngasse 3, 1090 Vienna, Austria}

\author{David Grass}
\thanks{U.D. and D.G. contributed equally to this work.}

\author{Manuel Reisenbauer}
\affiliation{Vienna Center for Quantum Science and Technology (VCQ), Faculty of Physics, University of Vienna, Boltzmanngasse 5, A-1090 Vienna, Austria}
\author{Tobias Damm}
\author{Martin Weitz}
\affiliation{Institute for Applied Physics, University of Bonn, Wegelerstr. 8, D-53115 Bonn, Germany}
\author{Nikolai Kiesel}
\affiliation{Vienna Center for Quantum Science and Technology (VCQ), Faculty of Physics, University of Vienna, Boltzmanngasse 5, A-1090 Vienna, Austria}
\author{Markus Aspelmeyer}
\affiliation{Vienna Center for Quantum Science and Technology (VCQ), Faculty of Physics, University of Vienna, Boltzmanngasse 5, A-1090 Vienna, Austria}
\affiliation{Institute for Quantum Optics and Quantum Information (IQOQI), Boltzmanngasse 3, 1090 Vienna, Austria}

\date{\today}

%


\begin{abstract}
We report dispersive coupling of an optically trapped silica nanoparticle ($143~$nm diameter) to the field of a driven Fabry-Perot cavity in high vacuum ($4.3\times 10^{-6}~$mbar). We demonstrate nanometer-level control in positioning the particle with respect to the intensity distribution of the cavity field, which allows access to linear, quadratic and tertiary optomechanical interactions in the resolved sideband regime. We determine all relevant coupling rates of the system, i.e. mechanical and optical losses as well as optomechanical interaction, and obtain a quantum cooperativity of $C_Q = 0.01$. Based on the presented performance the regime of strong cooperativity ($C_Q > 1$) is clearly within reach by further decreasing the mode volume of the cavity.             

\end{abstract}

\maketitle
%
%
%
%
%

\section{Introduction}
Cavity optomechanics enables optical quantum control of mechanical motion. Realized in a plethora of different platforms, it promises diverse applications ranging from quantum sensors to hybrid devices for quantum information processing, and it opens new ways to address fundamental questions in macroscopic quantum physics \cite{Kippenberg2008a,Aspelmeyer2012,Aspelmeyer2014a,BowenMilburn2018}. Current state of the art optomechanical systems include cryogenically cooled solid state devices coupled to superconducting microwave cavities or nanophotonic stuctures that routinely operate in the quantum regime. Examples range from motional ground state laser cooling \cite{Teufel2011, Chan2011} to the generation of quantum squeezed states \cite{Wollman2015,Pirkkalainen2015,Lecocq2015}, non-Gaussian states \cite{O'Connell2010a,Hong2017} and entangled states \cite{Palomaki2013,Riedinger2018,Marinkovic2018,Ockeloen-Korppi2018} of micro-and nanomechanical motion.  

Coupling the motion of a \emph{levitated} object to an optical cavity provides new possibilities. At ultra-high vacuum, levitation enables excellent isolation of the mechanical motion from the environment, enhanced inertial sensitivity \cite{Gieseler2013, Ranjit2015, Rodenburg16, Ranjit2016, Hebestreit2018} as well as quantum optomechanics at room temperature. Furthermore, optical micromanipulation techniques provide the means to control the potential landscape, which allows access to anharmonic potentials, e.g. \cite{Cizmar2011}. Switching the potential completely off allows free dynamics to be investigated with new approaches to force sensing and matter wave interferometry  \cite{Romero-Isart2011b,Kaltenbaek2012,Hebestreit2018}.

In its original form, levitated cavity optomechanics is realized by positioning an optically trapped particle inside an optical mode of a Fabry-Perot cavity. This configuration has first been suggested by \cite{Romero-Isart2010a,Chang2010, Barker2010} and builds on earlier work by \cite{Ritsch2001,Vuletic2000}. The presence of the particle shifts the cavity resonance frequency, which couples the particle motion dispersively to the cavity field, resembling the fundamental cavity optomechanical interaction. Cooling the nanoparticle motion is achieved by driving the cavity with a laser that is detuned to frequencies smaller than the cavity resonance frequency. Off-resonant (anti-Stokes) scattering will result in a cooling rate given by $\Gamma_{opt}\approx 4g^2/\kappa$ (with linear optomechanical coupling strength $g$ and the cavity linewidth $\kappa$), which competes with the heating rate $\Gamma$ due to the thermal environment. Cavity cooling to the ground state (as well as coherent quantum control in the resolved sideband regime) requires the ratio of these rates, the cooperativity $C_Q$, to exceed $1$. This condition, $C_Q=4g^2/\kappa \Gamma>1$, is called the strong cooperativity regime.

Early experiments have achieved optical trapping of nanoparticles in a Fabry-Perot cavity with a cooperativity of $C_Q \approx 2\times 10^{-6}$, limited by the environmental coupling at a gas pressure of 4 mbar \cite{Kiesel2013b}.  At high vacuum levels, coupling to an optical cavity has been achieved for particles that were not continuously localized in the cavity field, i.e. they were either in transit through the cavity \cite{Asenbaum2013} or confined to a shallow Paul trap \cite{Millen2015}. The largest dispersive shift up to now has been demonstrated for particles in transit through a microcavity \cite{Kuhn2017} and in plasmonic trapping \cite{Mestres2016}. Photonic microcavities also enabled significantly enlarged coupling for trapped particles \cite{Magrini2018}, but the cooperativity is still limited to $C_Q=10^{-9}$ due to environmental pressure.

Here we demonstrate orders of magnitude improvement in cooperativity for a levitated nanoparticle that is positioned inside a high-finesse Fabry-P\' erot cavity at high vacuum. We independently determine the optomechanical coupling rate $g$ from optomechanically induced transparency (OMIT) measurements, the mechanical heating rate $\Gamma$ via relaxation measurements \cite{Jain2016} and the optical losses $\kappa$ of the cavity. Based on these measurements we derive a quantum cooperativity \exCq  for our present system corresponding to an improvement by four orders of magnitude.

In this article, we first introduce the experimental setup, the technological developments necessary to combine a tweezer with a Fabry-P\'erot cavity and the loading procedure in chapter \ref{ch:setup}. A detailed analysis of the coupling between the particle motion and the cavity, and the particle mass is presented in chapter \ref{ch:character}. We proceed to characterize the mechanical heating rate $\Gamma$ with relaxation measurements in chapter \ref{ch:mechanicalLosses} and the optomechanical coupling rate $g$ using OMIT measurements in chapter \ref{ch:omit}. These results allow us to estimate the quantum cooperativity 
in chapter \ref{ch:conclusion}.

\section{Experimental Setup}
\label{ch:setup}
\begin{figure}[h]
	\centering
	\includegraphics{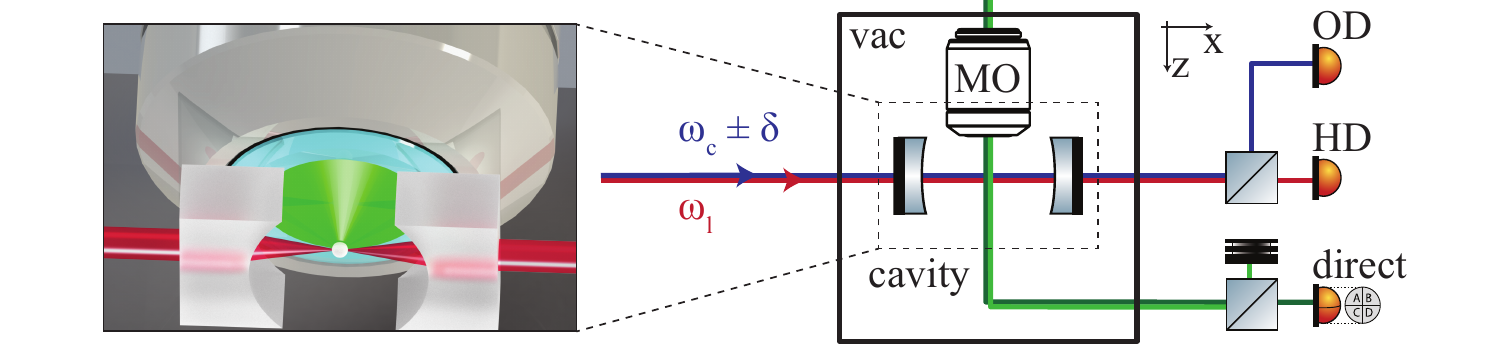}
	\caption{\label{fig:setup} Experimental Setup. A microscope objective (MO) and a trapping laser (dark green line) form an optical tweezer holding a silica nanoparticle in vacuum. The particles center of mass motion is coupled to a high-Finesse Fabry-Perot cavity which is inside the same vacuum chamber (vac). A feedback laser (light green line) in combination with a three-dimensional position read-out (read-out) is used to cool the particle motion by parametric feedback in the state-of-the-art scheme \cite{Gieseler2012b}. A laser (red line) is locked to the cavity resonance with frequency $\omcavi$. The transmission of this laser is used for homodyne-detection (HD) to monitor the particle motion along the cavity axis ($x$). We derive a control mode (blue line) from the locked laser beam at frequency $\omcavii\pm\delta$ whose detuning can be arbitrarily chosen with respect to the cavity resonance frequency. 
		For optomechanically induced transparency measurements the control mode is red-detuned with respect to the cavity resonance, phase modulated by a frequency $\delta$ and detected in transmission of the cavity (OD).  
		To give an impression of the geometry for holding the particle with a high-NA objective in the center of the optical cavity, the inset shows a artistic view of the cavity and the microscope objective.}
\end{figure}

The experiment combines a free space Fabry-P\'erot resonator with an optical tweezer as shown in Figure \ref{fig:setup}. The optical tweezer is formed by a laser at a wavelength of $\lambda=1064$ nm (light green line, trapping laser). We focus the light using a microscope objective (MO, Olympus LMPL100x IR) of long working distance ($\wodi=3.4$ mm) and a high numerical aperture ($\textrm{NA}=0.8$). The center-of-mass (COM) motion of a particle inside this Gaussian potential can be approximated by a three-dimensional harmonic oscillator with typical oscillation frequencies of $\omx=2\pi\times 170$ kHz and $\omy=2\pi\times190$ kHz along the radial tweezer directions ($x$ and $y$) and $\omz=2\pi\times 40$ kHz along the axial tweezer direction ($z$). The microscope objective is mounted on a triaxial nanopositioner such that the optical tweezer can be remotely positioned while a particle is in the trap. The particle COM motion is monitored in all three directions (standard detection, see \cite{Gieseler2012b}), using a collimation lens ($\textrm{NA}=0.16$, not shown) to collect the forward scattered light. To cool the COM motion in all directions we apply parametric feedback cooling, i.e. we remove energy from the particle motion by introducing a velocity dependent spring constant (for a detailed description, see \cite{Gieseler2012b}). This requires a modulation of the tweezer power at twice the mechanical frequency of the nanoparticle motion. We implement this modulation using a second laser (dark green line, feedback laser), which is coupled into the same spatial mode as the trapping laser. To avoid interference, the feedback laser is orthogonally polarized and shifted in frequency by 82 MHz with respect to the trapping laser. The modulation signal is generated from the nanoparticle position read-out using three phase-locked loops. This ensures that the particle is stably levitated in the optical tweezer over essentially indefinite times even in high vacuum. With feedback cooling we obtain an effective COM temperature of $T_{eff}=100$ mK in a room temperature environment at the base pressure of our vacuum system \exBasePressure.

The optomechanical interface is realized with a high-finesse ($\mathcal{F} \approx 73,000$), near-confocal Fabry-P\'erot cavity. More specifically we use a mirror with radius of curvature of $R=1$ cm, a cavity length of $L=1.07$ cm which results in a free spectral range of $\Delta\omega=2\pi\times 14.019$ GHz. The cavity is placed perpendicular with respect to the optical tweezer axis (along the $x$-axis, see figure \ref{fig:setup}). To maximize coupling, the optical tweezer needs to levitate the nanoparticle at the center of the optical cavity mode. Due to the high numerical aperture of the microscope objective this requires the cavity mirrors to be sufficiently narrow in $z$-direction to avoid clipping of the optical tweezer mode. The mirrors also need to be sufficiently wide to avoid clipping losses of the intracavity mode. 
To achieve this, we use mirrors with an original diameter of $12.7$~mm that are cut into $W_M=4$~mm wide strips (see Fig.~\ref{fig:setup}). The intracavity mode size on the cavity mirrors is \exCavityModeMirror$\ll W_M$ and clipping is negligible.

Two laser modes drive the Fabry-P\'erot cavity (Figure~\ref{fig:setup}).
The first mode (locking laser) is emitted from a laser ($\lambda=1064$ nm, frequency $\omcavi$) which is stabilized to the cavity using a Pound-Drever-Hall locking scheme. The light of the locking laser transmitted through the cavity is used for homodyne detection (HD) of the particle COM motion along the cavity axis. A second mode with frequency $\omcavii=\omcavi+\Delta\omega+\Delta$ (control laser) is derived from the locking laser: To this end, the phase is modulated with an electro-optical modulator (EOM) at frequency $\Delta\omega+\Delta$. One of the two created sidebands is selected with a filtering cavity. Thus, the control mode can be detuned by a variable frequency $\Delta$ with respect to the cavity mode that is one free spectral range $\Delta\omega$ away from the locking mode. This is sufficient for optomechanical control such as cavity cooling. In addition, we can apply a phase modulation with frequency $\delta$ to the control mode $\omcavii$ using the same EOM such that two sidebands are created at $\omcavii\pm\delta$. These are used for OMIT measurements as described in chapter \ref{ch:omit}. The control mode and the locking mode are orthogonally polarized and separated with a polarizing beamsplitter in transmission of the optomechanical cavity for homodyne detection (HD) and OMIT detection (OD), respectively.

With the experimental setup in place, the loading of silica nanoparticles (radius $a=(71.5\pm2)$~nm, Microparticles GmbH) into the tweezer is based on a nebulizer approach \cite{Burnham2006}. A medical nebulizer (Omron MicroAIR U22) is filled with a solution of silica nanoparticles and isopropanol in a mass ratio of 1:$10^4$. The nebulizer creates small droplets of liquid at ambient pressure that are sucked into the pre-evacuated vacuum chamber. Here a nanoparticle is eventually trapped at a pressure of typically $\sim 100$ mbar. While the density of nanoparticles we use allows for reproducible trapping in the optical tweezer, it turned out detrimental for the cavity mirrors. We found a reduction in finesse from $\mathcal F=200,000$ to $\mathcal{F}=40,000$ during a single loading attempt. To avoid contamination of the mirrors we designed our experiment in a modular fashion that allows removing only the optical cavity \cite{Grass2018}: The two cavity mirrors are glued into an aluminium mount which can be manually removed (inserted) from (into) the vacuum chamber via a CF quick access door. The cavity is not present during the loading procedure and only put back in place when a particle is levitated in the tweezer.

\section{Characterization of optomechanical coupling and particle mass}
\label{ch:character}
The setup presented above allows to change the position of a nanoparticle precisely with respect to the optical cavity. In the following we study the dependence of the optomechanical coupling on the particle position. It is convenient to separate the particle motion along the cavity axis $x=x_0 + \hat x$ into its average position $x_0$, the position of the potential minimum of the tweezer and its oscillatory motion $\hat x$ around $x_0$ (the same can be done for the remaining directions). The cavity frequency shift as function of the particle position is given by \cite{Chang2010,Nimmrichter2010}
\begin{equation}
\label{eq:cavityFreqShift}
U(x,y,z) 	= \frac{\omcavi\alpha}{2\varepsilon_0\vcav}\frac{w_0^2}{w^2(x_0)}e^{-\frac{y_0^2+z_0^2}{w^2(x_0)}}\sin^2(k x) = U_0\sin^2(k x)
\end{equation}
with the particle polarizability $\alpha=4\pi\varepsilon_0a^3(\varepsilon-1)/(\varepsilon+2)$, the vacuum permittivity $\varepsilon_0$, the dielectric constant of the particle $\varepsilon$, the cavity mode volume $\vcav=\pi w_0^2 L/4$, the cavity waist function $w(x)=w_0\sqrt{1+(x/x_R)^2}$ with a waist of \exCavityWaist and the Rayleigh length of the cavity mode $x_R$. Note that $U_0=U_0(x_0,y_0,z_0)$ is a function of the particle's average position. The particle position dependent change in resonance frequency of the cavity from equation \ref{eq:cavityFreqShift} gives rise to the optomechanical coupling. The corresponding interaction Hamiltonian is:
\begin{eqnarray}
\label{eq:interactionHam}
H_{int}	&=&	-\hbar \sqrt{\ncavi} U_0\sin^2\big[k(x_0+\hat x)\big] (\hat{a}_l^\dagger+\hat{a}_l)\nonumber\\
&\approx&	-\hbar\sqrt{\ncavi}\bigg[  g_0\frac{\hat x}{\xzpf} + g_q\frac{\hat{x}^2}{\xzpf^2} - g_c\frac{\hat{x}^3}{\xzpf^3}\bigg](\hat{a}_l^\dagger+\hat{a}_l)
\end{eqnarray}
with the intracavity cavity photon number $\ncavi$ and the zero-point fluctuation of the particle motion $\xzpf$. The approximation results from a Taylor expansion of the sinusoidal function around $x_0$ to third order and dropping the constant term. We also introduced the linear single photon coupling $g_0=\xzpf k U_0\sin(2kx_0)$, the quadratic coupling $g_q=\xzpf^2k^2U_0\cos(2kx_0)$ and the cubic coupling $g_c=\frac{2}{3}\xzpf^3k^3U_0\sin(2kx_0)$. 

The three terms in the interaction Hamiltonian lead to a modulation of the cavity susceptibility at the respective harmonics of the mechanical frequency. Experimentally, this is reflected in a phase modulation of the locking beam transmitted through the optical cavity and can be detected with homodyne detection. The amplitudes of the detected signals at their corresponding harmonics allow direct inference of linear, quadratic and cubic coupling. This also serves as alignment signal to positioning the optical tweezer with respect to the cavity mode.

\begin{figure}[h!]
	\centering
	\includegraphics{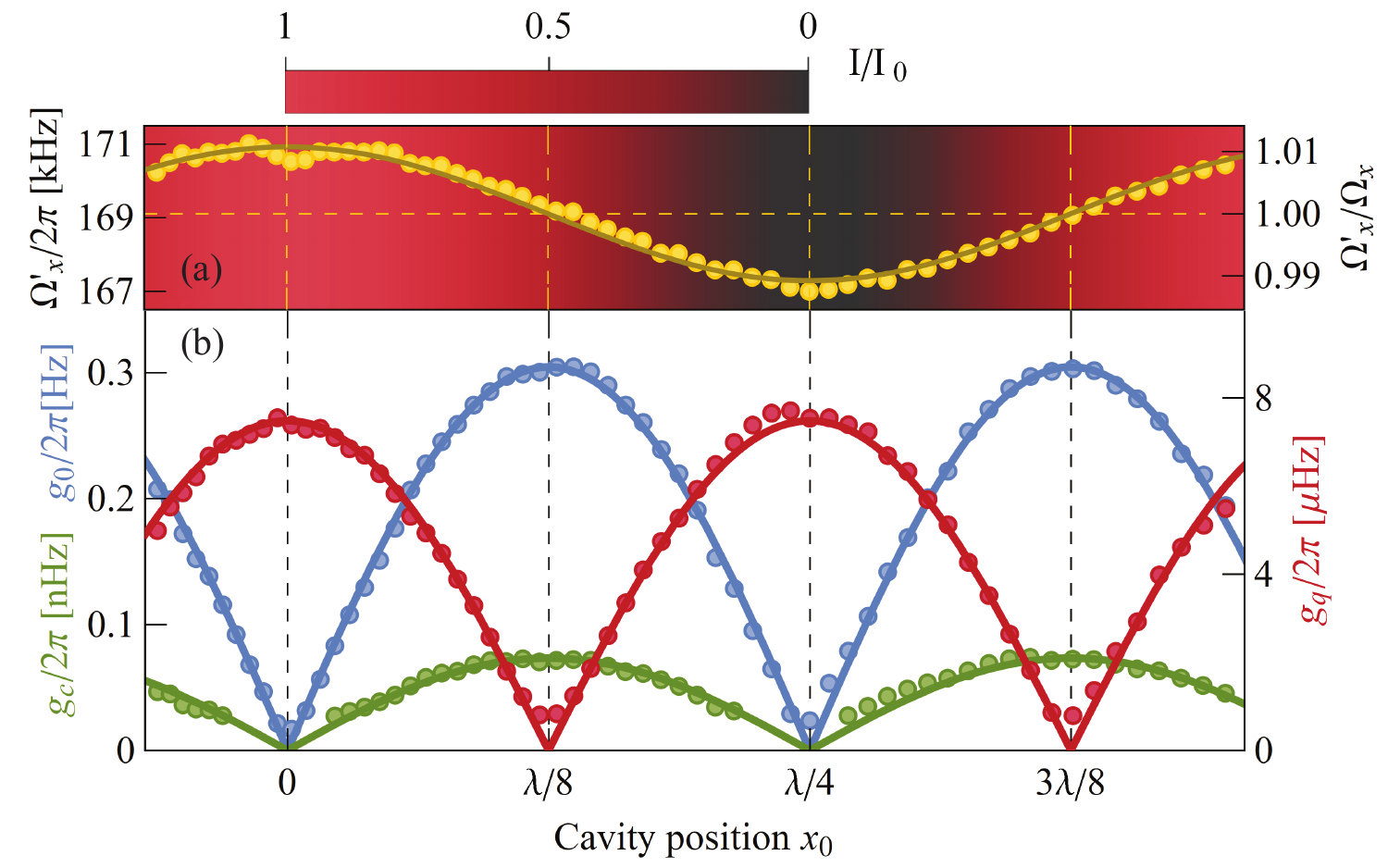}
	\caption{\label{fig:cavityscan} Position dependent frequency and optomechanical coupling. (a) The mechanical frequency $\Omega_x$ follows the intensity profile (background colour) as we move the nanoparticle along the cavity standing wave. The cavity adds a positive (negative) spring constant at the intensity maximum $x_0=0$ (minimum $x_0=\lambda/4$). The corresponding frequency variation is proportional to intracavity photon number of the locking mode. (b) Different orders of dispersive optomechanical coupling: The linear single photon coupling $g_0$ (blue) is maximal at the intensity slope ($x_0=\lambda/8$ and $x_0=3\lambda/8$) as well as the cubic coupling $g_c$ (green). Quadratic coupling $g_q$ (red) is optimal at the extrema of the intensity distribution.}
\end{figure}
In a first measurement we move the particle along the $y$-axis ($z$-axis) through the cavity mode, record a homodyne spectrum at each position, and extract the peak height $\propto U_0$ at the mechanical frequency $\omx$. The frequency shift introduced by the particle resembles the Gaussian envelope of the cavity mode, see Equation \ref{eq:cavityFreqShift}. Our measured value for the waist along the $y$-direction ($z$-direction) is \exWaistY  (\exWaistZ), which agrees well with the waist $w_0=41.1~\mu$m we expect from calculations based on radius of curvature and cavity length. In a second measurement we position the particle on the cavity axis ($y_0=z_0=0$), scan the levitated particle along the cavity axis $x$ in steps of 8 nm and record a homodyne spectrum at each position. 
We use the spectrum to map out the linear (blue), quadratic (red) and cubic (green) coupling strengths as a function of position (Figure \ref{fig:cavityscan} (b)). 

The amplitude of the nanoparticle motion enters the linear, the quadratic and the cubic coupling with different exponents, respectively, and can therefore be calculated explicitly from this measurement via the equipartition theorem. For example, the ratio between the second and first harmonic of the COM motion observed in  the homodyne detection is $g_q^2\langle x^4\rangle/(g_0^2x_{zpf}^2\langle x^2\rangle)=k^2 k_BT_0/(m\omx^2)$. As the environmental temperature $T_0$ and the frequency of the particle are precisely known, this allows to determine the mass of the nanoparticle:  $m=(2.86\pm 0.04)$~fg. 
Using the manufacturer specification for the density of silica $\rho=1850$ kg/m$^3$ and a spherical shape, we calculate a nanoparticle radius of $a=(71.8\pm 0.9)$ nm, in good agreement with the value of $a=(71.5\pm2)$ nm specified by the manufacturer. 

In addition, the locking field exerts a gradient force onto the nanoparticles modifying its optical potential. When the particle is moved along the cavity axis, this effect is observed in a position dependent modulation of the mechanical frequency $\Omega'_x=\omx+2g_qn_l\cos(2kx_0)$ as displayed in Figure \ref{fig:cavityscan}(a). The yellow dots represent data points, the solid line is a fit to the theory and the red-black shaded background indicates the intensity of the cavity field. At a high intensity region the additional confinement of the cavity field increases the mechanical frequency and vice versa at the cavity field node. Both measurement routines consistently determine the tweezer position with respect to the cavity field and are used to control the coupling between the particle COM motion and cavity field accordingly.

\section{Mechanical Heating rate $\Gamma$}
\label{ch:mechanicalLosses}
The optomechanical control competes with noise acting on the nanoparticles center-of-mass motion. The noise originates from two contributions: on one hand collisions with the surrounding gas molecules result in a heating rate  $\Gamma_p=\bar n_{th}\gamma_p$, with $\gamma_p$  the pressure dependent mechanical linewidth \cite{Beresnev1990} and $\bar{n}_{th}=\kb T_0/(\hbar\omx)$ the thermal occupation of the oscillator COM motion in equilibrium with the thermal environment. On the other hand, the nanoparticle scatters photons from the optical trap effectively creating a white and Gaussian but anisotropic force noise. The resulting recoil heating rate along the cavity axis is $\garc=I \sigma_{sc}/(\mathrm{ 5 k m c}\omx)$ with $k=2\pi/\lambda$, $I$ the intensity of the light field at the particle position, $\sigma_{sc}$ the scattering cross section of the particle and $c$ speed of light \cite{Jain2016}. 

The recoil heating is dominated by the trapping field of the tweezer but has, in principle, also a contribution from the control and the locking field of the cavity. In our current regime of operation, however, these contributions can be neglected as the intensity of the trapping laser is much higher than the intensity of both cavity modes. In addition, at the base pressure of our vacuum system (\exBasePressure) the pressure dependent heating rate $\Gamma_p\approx2\pi\times 21$~kHz still dominates over the recoil heating rate $\garc\approx 2\pi\times 7$~kHz.

\begin{figure}[h]
	\centering
	\includegraphics{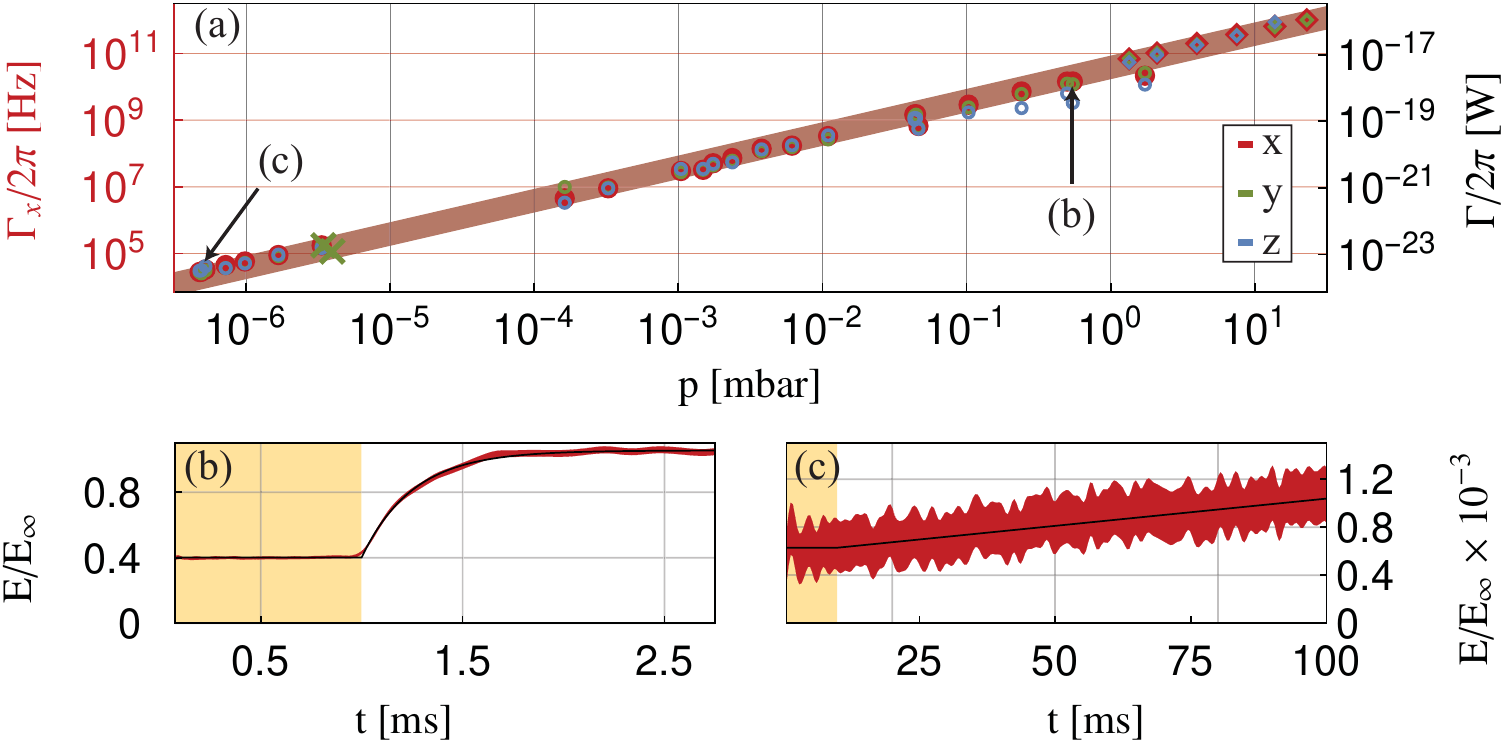}
	\caption{\label{fig:relaxation} Pressure dependent mechanical heating rate. (a) Mechanical heating rate $\Gamma_x/2\pi$ of a levitated particles center-of-mass motion along the cavity axis ($x$) as a function of pressure (red symbols). We measure the mechanical losses either via energy relaxation (circles and dots) at low pressures or via a spectral method (diamonds) at higher pressures. The phonon heating rate $\Gamma_x/2\pi$ (red) is shown for the motion in along the cavity (x-axis) as it depends on the direction of motion. The energy heating rate $\Gamma/2\pi$ $[W]$ (black) is expected equal in all spatial directions and applies to the data points for all directions of motion. The red shaded area is a theory curve whose width accounts for the uncertainties in system parameters (pressure, particle radius, mechanical frequency, equilibrium temperature) during the measurement. The two crosses mark heating rates with the cavity field on, showing that the cavity field has a negligible influence on heating. (b) and (c): Energy relaxation measurement: The particle is prepared in a low energy state with feedback cooling (yellow shaded area). After switching feedback cooling off, the particle relaxes towards thermal equilibrium. At a pressure of $p\approx0.56$ mbar (b), the full relaxation to thermal equilibrium is observed, while for lower pressures, i.e. $p=4\times 10^{-7}$ mbar (c) only the linear part of the relaxation was observed.}
\end{figure}

We measure the pressure dependent heating rate in the optical tweezer without the cavity fields present. We employ two complementary measurement protocols: Relaxation measurements analogous to \cite{Jain2016} in a pressure regime where feedback cooling is efficient ($p < 1$ mbar) and for higher pressures ($p>1$ mbar) we determine the linewidth of the mechanical oscillator from the noise power spectrum. In a relaxation measurement all three directions of motion are prepared in a low energy state  $E_0\ll E_\infty$ using parametric feedback cooling. After switching off the cooling, the mean particle energy evolves according to:
\begin{eqnarray}
\label{eq:relax}	
E(t)=	E_\infty + (E_0+E_\infty)e^{-\gamma t} \approx  E_0 + \Gamma t,
\end{eqnarray}	
with $\Gamma=\gamma E_\infty$, $\gamma=\gamma_p+\gamma_{rc}$ the combined mechanical linewidth, $\gamma_{rc}$ the recoil linewidth and $E_\infty$ the equilibrium energy (see supplementary information). Experimentally, we perform approximately 5000 repetitions of a relaxation measurement for each pressure with the particle initialized at a temperature much smaller than equilibrium with $E(t=0)=E_0$. Then, feedback cooling is switched off for a time $t_0$ over which the center-of-mass motion is allowed to thermalize. The energy is computed as an ensemble variance $E(t)\propto \langle x^2\rangle(t)$ of all 5000 relaxation trajectories. The resulting energy relaxation curves are shown for two cases in figure \ref{fig:relaxation}. In  part (b) the complete relaxation is monitored, which is possible as the particle can be levitated without feedback control at the corresponding pressure of $p=0.56$ mbar.  This is not possible at the base pressure of the vacuum system at \exBasePressure. Hence, we switch feedback cooling off only for a short period of time and resolve only the linear part of the relaxation trajectory. The solid black line is a fit of the energy data (red area) to equation \ref{eq:relax} with $\gamma,E_0$ and $E_\infty$ as free fit parameters. The yellow area indicates times when the feedback cooling is enabled. We infer the mechanical heating rates $\Gamma=\Gamma_p+\garc$ either directly from the linear fit or, when full relaxation is possible, from the product $\Gamma=\gamma E_\infty$. For pressures above 1~mbar we fit the mechanical linewidth $\gamma$ of a mechanical noise power spectrum and compute the heating rate via $\Gamma=\gamma E_\infty$ (see supplementary information). Figure \ref{fig:relaxation} (a) summarizes the mechanical loss measurements as a function of pressure. The COM motion along the cavity axis ($x$) is plotted in red (corresponding to the left $y$-scale: $\Gamma_x/2\pi$ [kHz]). The red shaded area is a theory plot accounting for all measurement uncertainties (pressure, particle radius, mechanical frequency, equilibrium temperature). The mechanical heating rate along the $y$ direction ($z$ direction) is plotted in green (blue) color. Together with the $x$ direction they share the right $y$-scale: $\Gamma /2\pi$ in units of Watts. The circles and dots represent relaxation measurements and the diamonds spectral measurements. As expected, in the air collision dominated regime $\Gamma_p\gg\garc$, the regime we are currently operating at, the mechanical losses are isotropic. 
To test the assumption that the cavity field has a negligible impact on the heating rate, we perform two relaxation measurements along the $y$-axis while the two cavity fields are present (green crosses). As expected, this causes no significant increase in heating rate compared to the case without cavity fields. At the base pressure of the vacuum system we measure the minimal mechanical heating rate along the cavity axis of $\Gamma=2\pi\times 28$ kHz. We expect to reach the recoil limit at a pressure of $p \approx 10^{-7}$ mbar.

\section{Optomechanically induced transparency}
\label{ch:omit}
To characterize the cooperativity $C_Q$ of our system we need to measure the coupling $g$ between optics and mechanics, for which we rely on optomechanically induced transparency (OMIT, \cite{Safavi-Naeini2011,Weis2010}). The phenomenon is analogue to electronically induced transparency in atom quantum optics \cite{Fleischhauer2005}. OMIT results from a modification of the cavity response in the presence of optomechanical interaction. More specifically, when the cavity is driven by a control laser (red-detuned from the cavity resonance by the mechanical frequency), near resonant light from a probe laser cannot enter the cavity. This effect is best understood in a scattering picture. The optomechanical interaction between particle and control mode scatters photons into the Stokes (far off-resonant) and anti-Stokes (resonant) sidebands of the control laser.  Destructive interference between the anti-Stokes photons and the probe laser photons in the cavity causes a reduced transmission of probe laser photons. The reduction in transmission of the probe laser is determined by the number of scattered photons, which only depends on the optomechanical coupling and not on the mechanical linewidth.

\begin{figure}[h]
	\centering
	\includegraphics{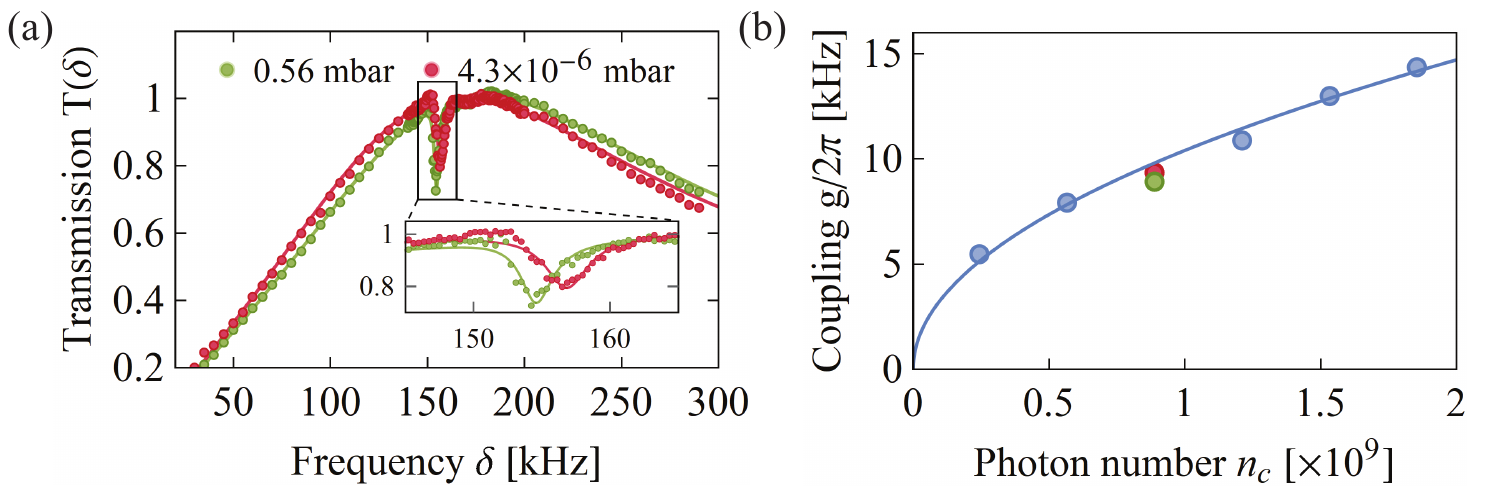}
	\caption{\label{fig:omit} Optomechanically induced transparency (OMIT) and optomechanical coupling rate $g$. (a) OMIT measurements: The cavity transmission of the probe beam is measured (OD) as a function of the probe modulation frequency $\delta$ 
		at a pressure of $p=0.56$ mbar (green) and $p= 4.3\times 10^{-6}$ mbar (red). 
		The envelope of the response is defined by the cavity linewidth $\kappa$ and the red-detuning of the control mode $\Delta$. The narrow dip is present at $\delta=\omx\approx 2\pi\times 160$ kHz due to destructive interference between nanoparticle motion and probe mode. It allows us to infer the optomechanical coupling rate $g$. 
		Inset: Zoom-in of the transparency window. The two curves differ due to a drift in detuning, which has no effect on the coupling rate. (b) Optomechanical coupling rate $g=g_0\sqrt{n_c}$ as a function of intracavity photon number $n_c$. Blue and green data points correspond to measurements at a pressure of $p=0.56$ mbar and the red data point corresponds to a measurement at $p=4.3\times 10^{-6}$ mbar demonstrating that the coupling is pressure independent. Note that the color coding in part (a) is reflected in the coupling plotted in part (b).
	}
\end{figure}

For our measurement, we first position the nanoparticle to maximize the single photon coupling $g_0$ between the particle and the control mode. As the particle is close to the longitudinal center of the optical cavity $x\approx 0$ we can achieve this by maximizing the linear coupling $g_0$ between the COM motion and the locking laser (see chapter \ref{ch:setup}). Then we switch on the control laser whose power determines the intracavity photon number and therefore the optomechanical coupling $g=\sqrt\ncavii g_0$. The probe laser is derived from the control laser by sideband modulation. More specifically, we phase modulate the control laser at a frequency $\delta$ creating two sidebands. One of them is far off-resonant with respect to the cavity and strongly suppressed  while the other one is used to probe the transmission near the cavity resonance. 

Figure \ref{fig:omit}(a) shows the normalized cavity transmission of the probe beam (detected with OD) as a function of the modulation frequency $\delta$ for two different pressures in the vacuum chamber (green: $0.56$ mbar, red: $4.3\times 10^{-6}$~mbar), i.e., two dissipation rates. The transparency window of the probe laser is clearly visible at the cavity resonance, in our case as a narrow window of reduced transmission. For a quantitative analysis the Stokes sideband, both modulation sidebands and the cavity linewidth have to be taken into account. Therefore the coupling $g$ is derived from a fit to the probe transmission with three free parameters: the mechanical linewidth $\gamma$, the red-detuning $\Delta$ and the mechanical frequency $\om$ (see supplementary information).

We use this method to study the dependence of the optomechanical coupling rate on the intracavity photon number $g=\sqrt\ncavii g_0$ (Figure~\ref{fig:omit}(a)), which is controlled with the control laser power. The expected square root dependence is clearly observable. All measurements were performed at a pressure of $p=0.56~$ mbar except for the red data point ($p=4.3\times 10^{-6}~$ mbar). It follows the same dependence and clearly demonstrates that the coupling we determine is pressure independent. The two data points highlighted in green and red correspond to the measurements in Figure~\ref{fig:omit}(a).

\section{Conclusion}
\label{ch:conclusion}

We have demonstrated dispersive coupling of a levitated nanoparticle in high vacuum (\exPressureOmit) to a high-finesse optical cavity. Excellent control over the nanoparticle position with respect to the cavity allowed us to observe linear, quadratic and cubic optomechanical coupling. We have determined the heating rates of the nanoparticle center-of-mass motion and employed optomechanically induced transparency to measure its coupling to the cavity field. This constitutes a complete set of independent measurements to determine the optomechanical cooperativity. With the measured value for the optomechanical coupling of \exCoupling, the mechanical heating rate \exGammaCq~ at a pressure of \exPressureOmit, and the optical losses \exKappa~ we obtain a quantum cooperativity of \exCq. This is the highest reported value for a dispersively coupled levitated nanoparticle so far (note the recent results on coherent scattering as an alternative approach \cite{Delic2018,Windey2018}). This is a major step towards the regime of strong quantum cooperativity ($C_Q>1$) in room temperature optomechanical systems. The straightforward path toward it would be to implement cavities with smaller mode volume \cite{RomeroLH}. We believe our experimental toolbox provides an important contribution to a future realization of quantum protocols that have been envisioned for levitated nanoparticles throughout the last decade, like quantum state preparation and matter-wave interferometry for tests of macroscopic quantum physics.


\bibliographystyle{apsrev4-1}
\bibliography{libary2}

\end{document}